\documentclass[12pt]{article}
\usepackage{amsmath,amsfonts}
\makeatletter \@addtoreset{equation}{section}

\usepackage{cite}
\usepackage{bm}
\usepackage{dcolumn}
\makeatother
\def\one{{\hbox{ 1\kern-.8mm l}}}
\newcommand{\Dslash}{\not{\hbox{\kern-4pt $D$}}}
\newcommand{\pdslash}{\not{\hbox{\kern-2pt $\partial$}}}
\newcommand{\be}{\begin{equation}}
\newcommand{\bea}{\begin{eqnarray}}
\newcommand{\eea}{\end{eqnarray}}
\newcommand{\ba}{\begin{array}}
\newcommand{\ea}{\end{array}}
\newcommand{\ee}{\end{equation}}

\usepackage{color}
\begin{document}

\begin{titlepage}
\vspace{10mm}

\vspace*{20mm}
\begin{center}
{\Large {\bf On the Butterfly Effect in 3D Gravity }\\
}
\vspace*{15mm}
\vspace*{1mm}
{Mohammad M. Qaemmaqami}

\vspace*{1cm}

{ \it  School of Particles and Accelerators\\Institute for Research in Fundamental Sciences (IPM)\\
	P.O. Box 19395-5531, Tehran, Iran }

\vspace*{0.5cm}
{E-mail: m.qaemmaqami@ipm.ir}%

\vspace*{2cm}
%%\maketitle

\vspace*{1cm}

\vspace*{0.5cm}

\vspace*{2cm}
%%\maketitle

\end{center}

\begin{abstract}
We study the butterfly effect by considering shock wave solutions near the horizon of the AdS black hole in some of 3-dimensional Gravity models including; 3D Einstein Gravity, Minimal Massive 3D Gravity, New Massive Gravity, Generalized Massive Gravity, Born-Infeld 3D Gravity and New Bi-Gravity. We calculate the butterfly velocities of these models and also we consider the critical points and different limits in some of these models. By studying the butterfly effect in the Generalized Massive Gravity, we observe a correspondence between the butterfly velocities and right-left moving degrees of freedom or the central charges of the dual 2D Conformal Field Theories.

\end{abstract}

\end{titlepage}

\section{Introduction}
Since it was shown\cite{Shenker:2013pqa},\cite{Shenker:2013yza},\cite{Roberts:2014isa},\cite{Leichenauer:2014nxa} that chaos in thermal CFT may be described by
shock wave near horizon of an AdS black hole, in other words, holographically the propagation of the shock
wave on the horizon provides a description of butterfly effect in the dual field theory. In field theory side butterfly effect may be diagnosed by out-of-time order four point function between pairs of local operators
\begin{eqnarray}
\langle V_{x}(0)W_{y}(t)V_{x}(0)W_{y}(t)\rangle_{\beta},
\end{eqnarray}
where $ \beta $ is inverse of the temperature. The butterfly effect may be seen by a sudden decay after the scrambling time, $ t_{*} $ ,
\begin{eqnarray}
\frac{\langle V_{x}(0)W_{y}(t)V_{x}(0)W_{y}(t)\rangle_{\beta}}{\langle V_{x}(0)V_{x}(0)\rangle_{\beta} \langle W_{y}(t)W_{y}(t)\rangle_{\beta}}\sim 1-e^{\lambda_{L}\big(t-t_{*}-\frac{|x-y|}{v_{B}}\big)},
\end{eqnarray}
where $ \lambda_{L} $ is the Lyapunov exponent and $ v_{B} $ is the butterfly velocity. The Lyapunov exponent is, $ \lambda_{L}=\frac{2\pi}{\beta} $, where $ \beta $ is inverse of Hawking temperature. And also the butterfly velocity should be identified by the velocity of shock wave by which the perturbation spreads in the space. People have done some works recently on butterfly effect and it's different aspects\cite{Alishahiha:2016cjk},\cite{Blake:2016wvh},\cite{Ling:2016ibq},\cite{Ling:2016wuy},\cite{Feng:2017wvc},\cite{Kim:2017dgz},\cite{Cai:2017ihd},\cite{Baggioli:2016pia},\cite{Baggioli:2017ojd},\cite{Patel:2017vfp},\cite{Blake:2017qgd},\cite{Qi:2017ttv},\cite{Ling:2017jik} and also it needs more investigation and calculation to access a better understanding of this interesting natural phenomenon.\\
\\
Gravity in 3-dimensions is special and interesting, because it's a good toy model for the quantum gravity and by studying 3D Gravity we can access a deeper understanding about gravity in higher dimensions, and also is a good context for holography because we are more familiar with conformal field theories in 2-dimensions in comparison with other dimensions. In addition we can construct the ghost free higher derivative gravity models in 3-dimensions. The 3D Einstein Gravity do not have propagating degrees of freedom or gravitons in the bulk but by adding higher derivative terms to the action we can have massive propagating degrees of freedom. For example by adding gravitational Chern-Simons action to the 3D Einstein Gravity action we have a massive graviton in the linearized level of the Topologically Massive Gravity theory\cite{Deser:1981wh}. \\
\\
In this paper we study the butterfly effect in some of 3D Gravity models, we calculate the butterfly velocities of these models and also consider the critical points and different limits in some of them. In section 2 we study the butterfly effect in the 3D Einstein Gravity and we find that the butterfly velocity of 3D Einstein Gravity is equal to velocity of light. In section 3 we study the butterfly effect in the Minimal Massive 3D Gravity\cite{Bergshoeff:2014pca} which is proposed for resolving the bulk-boundary clash problem in the Topologically massive Gravity(TMG)\cite{Deser:1981wh} and we consider the TMG limit and the critical point in this model.\\
\\
In section 4 we first review the butterfly effect in the New Massive Gravity\cite{Bergshoeff:2009hq} by details then we study the Generalized Massive Gravity\cite{Bergshoeff:2009hq},\cite{Liu:2009pha} and it's different limits and the critical lines. then we observe a correspondence between the butterfly velocities and the central charges of the dual 2D Conformal Field Theory. In section 5 we study the butterfly effect in the Born-Infeld 3D Gravity \cite{Gullu:2010pc}, \cite{Gullu:2010st}, \cite{Alishahiha:2010iq} and it's critical point and we see at the critical point of the model, both of the butterfly velocities vanish. In section 6 we study the butterfly effect in the New Bi-Gravity\cite{Akhavan:2016hju},\cite{Akhavan:2016xmj} and consider the causality bound in this model and also we consider the logarithmic solutions limit of the New Bi-Gravity. The last section is devoted to conclusions.
\section{3D Einstein Gravity}
The action of the 3D Einstein Gravity with a cosmological constant is:
\begin{eqnarray}
\frac{1}{16\pi G}\int d^{3}x\sqrt{-g}(R-2\Lambda).
\end{eqnarray}
If we vary the action with respect to metric we find the equations of motion as follows,
\begin{eqnarray}
R_{\mu\nu}-\frac{1}{2}Rg_{\mu\nu}+\Lambda g_{\mu\nu}=\kappa T_{\mu\nu}.
\end{eqnarray}
To study the butterfly effect, we must to consider the black hole solution. the equations of motion of the 3D Einstein Gravity admit this asymptotically AdS black hole solution similar to non rotating BTZ black hole \cite{Banados:1992wn}, \cite{Banados:1992gq}:
\begin{eqnarray}
ds^{2}=-f(r)dt^{2}+\frac{dr^{2}}{f(r)}+r^{2}d\varphi^{2},    \,\,\,\,\    \,\,\,\,\    f(r)=\frac{r^{2}}{l^{2}}\bigg(1-\frac{r_{h}^{2}}{r^{2}}\bigg), \,\,\,\,\      \,\,\,\,\    \Lambda= -\frac{1}{l^2},
\end{eqnarray}
where $ r_{h} $ is the radius of horizon and $ l $ is AdS radius. The $ \varphi $ coordinate is dimensionless and compact; $ 0\leq\varphi \leq 2\pi $, now let us introduce a coordinate with length dimension $ x= l\varphi $, then we have $ d\varphi= \frac{dx}{l} $, and also $ x $ coordinate is compact, $ 0\leq x\leq 2\pi l $, therefore the AdS black hole metric is:
\begin{eqnarray}
ds^{2}=-f(r)dt^{2}+\frac{dr^{2}}{f(r)}+\frac{r^{2}}{l^{2}} dx^{2}.
\end{eqnarray}
Now the aim is to study the shock wave of this model when the above black hole solution of this theory is perturbed by injection of a small amount of energy. For this aim, it is better to rewrite the solution in the Kruskal coordinate\cite{Shenker:2013pqa},
\begin{eqnarray}
u=exp\big[\frac{2\pi}{\beta}(r_{*}-t)\big],  \,\,\,\,\    \,\,\,\,\   v=-exp\big[\frac{2\pi}{\beta}(r_{*}+t)\big],
\end{eqnarray}
where $ \beta=\frac{4\pi}{f^{'}(r)} $ is the inverse of temperature and $ dr_{*}=\frac{dr}{f(r)} $ is the tortoise coordinate.\\
\\
By making use of this coordinate system, the metric becomes into this form\cite{Shenker:2013pqa},\cite{Alishahiha:2016cjk}:
\begin{eqnarray}
ds^{2}=2A(uv)dudv+B(uv)dx^{2}.
\end{eqnarray}
Here $ A(uv) $ and $ B(uv) $ are two functions, given by $ f(r) $, whose near horizon expansions are
\begin{eqnarray}
A(uv)&=&-2cl^{2}\bigg(1-2cuv+3c^{2}u^{2}v^{2}-4c^{3}u^{3}v^{3}+...\bigg),\\\nonumber
B(uv)&=&\frac{r_{h}^{2}}{l^{2}}\bigg(1-4cuv+8c^{2}u^{2}v^{2}-12c^{3}u^{3}v^{3}+...\bigg),
\end{eqnarray}
where $ c $ is an integration constant to be fixed later. Now we must study the shock wave, for this aim let us consider an injection of a small amount of energy from boundary toward the horizon at time $ -t_{w} $. This will cross the $ t=0 $ time slice while it is red shifted. Therefore the equations of motion should be deformed as 
\begin{eqnarray}
\mathcal{E}_{\mu\nu}=\kappa T^{s}_{\mu\nu},
\end{eqnarray}
where $ \kappa=8\pi G_{N} $, the energy-momentum tensor has only $ uu $ component due to energy injection:
\begin{eqnarray}
T_{uu}^{S}=lE\bigg(exp\big(\frac{2\pi t_{w}}{\beta}\big)\delta(u)\delta(x)\bigg).
\end{eqnarray}
For solving the equations of motion near horizon to find the shock wave solution, we consider this ansatz for back-reacted geometry
\begin{eqnarray}
ds^{2}=2A(UV)dUdV+B(UV)dx^{2}-2A(UV)h(x)\delta(U)dU^{2},
\end{eqnarray} 
where the new coordinate $ U $ and $ V $ are
\begin{eqnarray}
U\equiv u,   \,\,\,\,\     \,\,\,\,\  V\equiv v+h(x)\Theta(u).
\end{eqnarray}
Plugging the ansatz into the equations of motion Eq(2.2), near horizon at the leading order one finds a second order differential equation for $ h(x) $
\begin{eqnarray}
\big(l^{4}\partial_{x}^{2}-r_{h}^{2}\big)h(x)=-\frac{r_{h}^{2}}{2c}\big(\kappa lEe^{2\pi t_{w}/\beta}\big)\delta(x),
\end{eqnarray}
we can reduce the equation of motion into:
\begin{eqnarray}
(\partial_{x}^{2}-a^{2})h(x)=\xi\delta(x),   \,\,\,\,\,\,\,\,\,\      a^{2}=\frac{r_{h}^{2}}{l^{4}},   \,\,\,\,\,\,\,\,\,\      \xi=-\frac{r_{h}^{2}}{2cl^{4}}\big(\kappa lEe^{2\pi t_{w}/\beta}\big),
\end{eqnarray}
 whose solution is
 \begin{eqnarray}
 h(x)=-\frac{\xi}{2a}e^{-a|x|}.
 \end{eqnarray}
 By replacing the values of $ a $ and $ \xi $, one can see, $ h(x)\sim e^{\frac{2\pi}{\beta}\big[(t_{w}-t_{*})-|x|/v_{B}\big]} $, where the scrambling time is $ t_{*}=\frac{\beta}{2\pi}log(\frac{l}{\kappa}) $, with $ \kappa=8\pi G_{N} $ and $ G_{N} $ is Newton's constant in $ D=3 $, for true value of scrambling time\cite{Roberts:2014isa}, if we assume; $ lE\sim 1 $, we have to fix the value of integration constant to $ c=\frac{r_{h}}{4l} $, in the above expression. Then one can read the value of the butterfly velocity\cite{Roberts:2014isa},\cite{Alishahiha:2016cjk},\cite{Qaemmaqami:2017bdn}:
 \begin{eqnarray}
 v_{B}=\frac{2\pi}{\beta a}=1,   \,\,\,\,\    \,\,\,\,\    \frac{2\pi}{\beta}=\frac{f^{'}(r)}{2}=\frac{r_{h}}{l^{2}}.
 \end{eqnarray}
which is in agreement with \cite{Shenker:2013pqa}, where the butterfly velocity in the Einstein Gravity in D-Dimension is $ v_{B}= \sqrt{\frac{D-1}{2(D-2)}} $. Note that the largest possible butterfly velocity in the Einstein Gravity is in $ D=3 $ which is equal to light velocity($ v_{B}=1 $). It is important to note, although in the 3D Einstein Gravity there is no propagating degrees of freedom in the bulk, due to boundary degrees of freedom or boundary gravitons the butterfly velocity is non-zero. It's a sign of relationship between butterfly velocities and boundary degrees of freedom or boundary gravitons. Note that in contradiction of no propagating degrees of freedom in the bulk in 3D Einstein Gravity, it's dual 2D conformal field theory has non-zero central charge\cite{Brown:1986nw}, in the forth section we will see a correspondence between butterfly velocities and central charges of dual 2D conformal field theory.
\section{Minimal Massive 3D Gravity}
The Minimal Massive 3D Gravity(MMG) is a model which is proposed for resolving bulk-boundary clash problem in the Topologically Massive Gravity(TMG) (we have not positive energy of graviton and unitary dual 2D conformal field theory at the same time in TMG.)\cite{Deser:1981wh}, with adding a new term to the action in the vielbein formalism \cite{Bergshoeff:2014pca}. And also we know that the linearized equations of motion of MMG is equal to linearized equations of motion of TMG by making use a redefinition of topological mass parameter\cite{Tekin:2014jna},\cite{Alishahiha:2014dma},\cite{Alishahiha:2015whv}, therefor the model has a single local degree of freedom that is realized as a massive graviton in linearization as TMG. The Lagrangian of the Minimal Massive 3D Gravity in vielbein formalism is:
\begin{eqnarray}
L_{MMG}= -\sigma e.R+\frac{\Lambda_{0}}{6}e.e\times e+h.T(\omega)+\frac{1}{2\mu}\big(\omega.d\omega+\frac{1}{3}\omega.\omega\times\omega\big)+\frac{\alpha}{2}e.h\times h,
\end{eqnarray}
where $ e $, is vielbein, $ \omega $ is spin connection and $ h $ is a Lagrange multiplier or auxiliary field. Note that dot and cross mean internal and external product respectively, dot implies contraction of Lorenz indices of two fields with each other and cross means contraction of Lorenz indices of two fields with two indices of Levi-Chivita tensor. The equations of motion of MMG in metric formalism is:
\begin{eqnarray}
\bar{\sigma}G_{\mu\nu}+\bar{\Lambda}_{0}g_{\mu\nu} +\frac{1}{\mu}C_{\mu\nu}+\frac{\gamma}{\mu^{2}}J_{\mu\nu}=\kappa T_{\mu\nu},
\end{eqnarray}
where $ G_{\mu\nu} $ is Einstein tensor, $ C_{\mu\nu} $ is 3D Cotton tensor 
\begin{eqnarray}
C_{\mu\nu}= \epsilon_{\mu}~^{\alpha\beta}\nabla_{\alpha}\bigg(R_{\mu\nu}-\frac{1}{4}Rg_{\mu\nu}\bigg),
\end{eqnarray}
and $ J_{\mu\nu} $ is a curvature squared, symmetric tensor:
\begin{eqnarray}
J_{\mu\nu}=R_{\mu}~^{\lambda}R_{\lambda\nu}-\frac{3}{4}RR_{\mu\nu}-\frac{1}{2}g_{\mu\nu}\bigg(R_{\rho\sigma}R^{\rho\sigma}-\frac{5}{8}R^{2}\bigg).
\end{eqnarray}
Note that the relations between parameters of vielbein and metric formalisms are:
\begin{eqnarray}
\gamma = \frac{\alpha}{(1+\sigma\alpha)^2},~~\bar{\sigma} = - \left(\sigma + \alpha+\frac{\alpha^2\Lambda_{0}}{2\mu^{2}(1+\sigma\alpha)^{2}}\right),~~
\bar{\Lambda}_{0} = -\Lambda_{0} \left(1+\sigma\alpha-\frac{\alpha^{3}\Lambda_{0}}{4\mu^{2}(1+\sigma\alpha)^{2}}\right)\cr\nonumber\\
\end{eqnarray}
To study the butterfly effect, we must to consider the black hole solution. the equations of motion of the Minimal Massive 3D Gravity admit this asymptotically AdS black hole solution Eq(2.4) with:
\begin{eqnarray}
\bar{\Lambda}_{0}=-\frac{\gamma+4l^{2}\mu^{2}\bar{\sigma}}{4l^{4}\mu^{2}}.
\end{eqnarray}
Plugging the ansatz in the Kruskal coordinate Eq(2.10) into equations of motion Eq(3.2), near horizon at the leading order one finds a third order differential equation for $ h(x) $ 
\begin{eqnarray}
&&\frac{d^{3}h(x)}{dx^{3}}+\frac{r_{h}}{2\mu l^{3}}\Big(\gamma+2\mu^{2}l^{2}\bar{\sigma}\Big)h^{''}(x)-\frac{r_{h}^{2}}{l^{4}}h^{'}(x)-\cr\nonumber\\
&&-\frac{r_{h}^{3}}{2\mu l^{7}}\Big(\gamma+2\mu^{2}l^{2}\bar{\sigma}\Big)h(x)=-\frac{r_{h}^{3}\mu}{2cl^{5}}\big(\kappa lEe^{2\pi t_{w}/\beta}\big)\delta(x),
\end{eqnarray}
we can reduce the differential equation to:
\begin{eqnarray}
\big(\partial_{x}+a\big)\big(\partial_{x}^{2}-b^{2}\big)h(x)&=&\xi\delta(x),   \,\,\,\,\,\         \,\,\,\,\,\   a=\frac{r_{h}}{2\mu l^{3}}\Big(\gamma+2\mu^{2}l^{2}\bar{\sigma}\Big),\cr\nonumber\\  b&=&\frac{r_{h}}{l^{2}},  \,\,\,\,\,\,\,\           \,\,\,\,\,\,\,\    \xi=-\frac{r_{h}^{3}\mu}{2cl^{5}}\big(\kappa lEe^{2\pi t_{w}/\beta}\big),
\end{eqnarray}
we can decompose the above differential equation into two differential equation as follows:
\begin{eqnarray}
q^{'}(x)+aq(x)&=&\xi\delta(x),\cr\nonumber\\
h^{''}(x)-b^{2}h(x)&=&q(x),
\end{eqnarray}
the solution of first equation is
\begin{eqnarray}
q(x)=\xi\Theta(x)e^{-ax}.
\end{eqnarray}
If we solve the second equation by making use the above $ q(x) $ we find:
\begin{eqnarray}
h(x)=-\frac{\xi}{2b}\bigg(\frac{e^{-bx}}{a-b}-\frac{2be^{-ax}}{a^{2}-b^{2}}\bigg),
\end{eqnarray}
By replacing the values of $ a $, $ b $ and $ \xi $ we can read the scrambling time and the butterfly velocities\cite{Alishahiha:2016cjk},\cite{Alishahiha:2017hwg} as follows:
\begin{eqnarray}
t_{*}=\frac{\beta}{2\pi}log\frac{l}{\kappa},  \,\,\,\,\      \,\,\,\,\  v_{B}^{(1)}=\frac{2\pi}{\beta b}=1,  \,\,\,\,\   \,\,\,\,\  v_{B}^{(2)}=\frac{2\pi}{\beta a}= \frac{2\mu l}{\gamma+2\mu^{2}l^{2}\bar{\sigma}}, \,\,\,\,\   \,\,\,\,\  \frac{2\pi}{\beta}=\frac{f^{'}(r)}{2}=\frac{r_{h}}{l^{2}}.
\end{eqnarray}
The butterfly effect in the Topologically Massive Gravity(TMG) has been studied in\cite{Alishahiha:2016cjk} and they found the butterfly velocities as follows:
\begin{eqnarray}
v_{B}^{(1)}=1,    \,\,\,\,\,\,\            \,\,\,\,\,\,\    v_{B}^{(2)}=\frac{1}{\mu l}.
\end{eqnarray}
One can see the butterfly velocities of MMG in TMG limit($ \gamma=0, \bar{\sigma}=1 $) is equal to the butterfly velocites of TMG Eq(3.12).\\
Now lets consider the critical point of MMG, $ \gamma=-2\mu l(\mu l\bar{\sigma}-1) $ where massive and massless modes degenerates and the model has logarithmic solutions\cite{Alishahiha:2014dma},\cite{Alishahiha:2015whv} at this point, one can see at the critical point both of velocities degenerate and are equal to the butterfly velocity of the 3D Einstein Gravity which is equal to velocity of light:
\begin{eqnarray}
v_{B}^{(1)}=v_{B}^{(2)}=1.
\end{eqnarray}
\section{New Massive Gravity and Generalized Massive\\ Gravity}
The butterfly velocities of  the New Massive Gravity(NMG) has been obtained in\cite{Alishahiha:2016cjk}, here we review it by more details. The action of NMG is\cite{Bergshoeff:2009hq}:
\begin{eqnarray}
S_{NMG}=\frac{1}{16\pi G}\int d^{3}x\sqrt{-g}\bigg[R-2\Lambda -\frac{1}{m^{2}}\big(R_{\mu\nu}R^{\mu\nu}-\frac{3}{8}R^{2}\big)\bigg].
\end{eqnarray}
One can obtain the equations of motion by varying the action with respect to metric:
\begin{eqnarray}
G_{\mu\nu}+\Lambda g_{\mu\nu}-\frac{1}{2m^2}K_{\mu\nu}=\kappa T_{\mu\nu},
\end{eqnarray}
where
\begin{eqnarray}
K_{\mu\nu}=2\Box R_{\mu\nu}-\frac{1}{2}\big(\nabla_{\mu}\nabla_{\nu}R+g_{\mu\nu}\Box R\big)-8R_{\mu}~^{\sigma}R_{\sigma\nu}+\frac{9}{2}RR_{\mu\nu}+\bigg(3R_{\alpha\beta}R^{\alpha\beta}-\frac{13}{8}R^{2}\bigg)g_{\mu\nu}
\end{eqnarray}
The equations of motion of the New Massive Gravity admit this asymptotically AdS black hole solution Eq(2.4) with:
\begin{eqnarray}
\Lambda=-\big(\frac{1}{l^{2}}+\frac{1}{4m^{2}l^{4}}\big).
\end{eqnarray}
Plugging the ansatz in the Kruskal coordinate Eq(2.10) into equations of motion Eq(4.2), near horizon at the leading order one finds a forth order differential equation for $ h(x) $\footnote{The shock wave solution in Minkowski space background for TMG (and NMG) is studied in \cite{Edelstein:2016nml}}:
\begin{eqnarray}
\frac{d^{4}h(x)}{dx_{4}}-\frac{r_{h}^{2}}{2l^{4}}\big(3+2m^{2}l^{2}\big)h^{''}(x)+\frac{r_{h}^{2}}{2l^{8}}\big(1+2m^{2}l^{2}\big)h(x)=\frac{r_{h}^{4}m^{2}}{2cl^{6}}\big(\kappa lEe^{2\pi t_{w}/\beta}\big)\delta(x),
\end{eqnarray}
we can reduce the above differential equation to:
\begin{eqnarray}
&&\big(\partial_{x}^{2}-b_{1}^{2}\big)\big(\partial_{x}^{2}-b_{2}^{2}\big)h(x)=\xi\delta(x),   \,\,\,\,\      \,\,\,\,\     b_{1}^{2}= \frac{r_{h}^{2}}{l^{4}},\cr\nonumber\\
&& b_{2}^{2}= \frac{r_{h}^{2}}{2l^{4}}\big(1+2m^{2}l^{2}\big), \,\,\,\,\     \,\,\,\,\   \xi=\frac{r_{h}^{4}m^{2}}{2cl^{6}}\big(\kappa lEe^{2\pi t_{w}/\beta}\big),
\end{eqnarray}
we can decompose the above differential equation into two differential equation as follows:
\begin{eqnarray}
q^{''}(x)-b_{1}^{2}q(x)&=&\xi\delta(x),\cr\nonumber\\
h^{''}(x)-b_{2}^{2}h(x)&=&q(x),
\end{eqnarray}
By solving the first equation we have:
\begin{eqnarray}
q(x)=-\frac{\xi}{2b_{1}}e^{-b_{1}|x|},
\end{eqnarray}
By replacing the above $ q(x) $ in the second equation and solving the equation one find:
\begin{eqnarray}
h(x)= \frac{\xi}{2b_{1}b_{2}(b_{1}^{2}-b_{2}^{2})}\bigg(b_{1}e^{-b_{2}x}-b_{2}e^{-b_{1}x}\bigg).
\end{eqnarray}
Using the expressions for $ b_{1} $, $ b_{2} $ and $ \xi $ we can read the scrambling time and the butterfly velocities as follows:
\begin{eqnarray}
t_{*}=\frac{\beta}{2\pi}log\frac{l}{\kappa},  \,\,\,\,\      \,\,\,\,\  v_{B}^{(1)}=\frac{2\pi}{\beta b_{1}}=1,  \,\,\,\,\   \,\,\,\,\  v_{B}^{(2)}=\frac{2\pi}{\beta b_{2}}= \frac{1}{\sqrt{m^{2}l^{2}+\frac{1}{2}}}, \,\,\,\,\   \,\,\,\,\  \frac{2\pi}{\beta}=\frac{f^{'}(r)}{2}=\frac{r_{h}}{l^{2}}.\cr\nonumber\\
\end{eqnarray}
One can see at the critical point of NMG, $ m^{2}l^{2}=\frac{1}{2} $ \cite{Alishahiha:2010bw}, the two butterfly velocities degenerate into one velocity which is velocity of light
\begin{eqnarray}
m^{2}l^{2}=\frac{1}{2},  \,\,\,\,\            \,\,\,\,\    v_{B}^{(1)}=v_{B}^{(2)}=1.
\end{eqnarray}
Now lets consider the Generalized Massive Gravity(GMG) which is the combination of TMG and NMG\cite{Bergshoeff:2009hq},\cite{Liu:2009pha}, the action of GMG is action of NMG plus the gravitational Chern-Simons action:
\begin{eqnarray}
S_{GMG}=\frac{1}{16\pi G}\int d^{3}x\sqrt{-g}\bigg[R-2\Lambda -\frac{1}{m^{2}}\big(R_{\mu\nu}R^{\mu\nu}-\frac{3}{8}R^{2}\big)\bigg]+S_{CS},
\end{eqnarray}
where the gravitational Chern-Simons action is\cite{Deser:2002iw}:
\begin{eqnarray}
S_{CS}= \frac{1}{32\pi G\mu}\int d^{3}x\sqrt{-g}\epsilon^{\lambda\mu\nu}\Gamma^{\rho}_{\lambda\sigma}\big[\partial_{\mu}\Gamma^{\sigma}_{\rho\nu}+\frac{2}{3}\Gamma^{\sigma}_{\mu\tau}\Gamma^{\tau}_{\nu\rho}\big].
\end{eqnarray}
One can obtain the equations of motion by varying the action with respect to metric:
\begin{eqnarray}
G_{\mu\nu}+\Lambda g_{\mu\nu}-\frac{1}{2m^2}K_{\mu\nu}+\frac{1}{\mu}C_{\mu\nu}=\kappa T_{\mu\nu},
\end{eqnarray}
where $ K_{\mu\nu} $ is defined by Eq(4.3) and  $ C_{\mu\nu} $ is 3D Cotton tensor which is defined by Eq(3.3)
The equations of motion of the Generalized Massive Gravity admit this asymptotically AdS black hole solution Eq(2.4) with:
\begin{eqnarray}
\Lambda=-\big(\frac{1}{l^{2}}+\frac{1}{4m^{2}l^{4}}\big).
\end{eqnarray}
Plugging the ansatz in the Kruskal coordinate Eq(2.10) into equations of motion Eq(4.14), near horizon at the leading order one finds a forth order differential equation for $ h(x) $:
\begin{eqnarray}
&&\frac{d^{4}h(x)}{dx^{4}}-\frac{r_{h}m^{2}}{l\mu}\frac{d^{3}h(x)}{dx^{3}}-\frac{r_{h}^{2}m^{2}}{2l^{2}}\big(2+\frac{3}{m^{2}l^{2}}\big)h^{''}(x)+\frac{r_{h}^{3}m^{2}}{\mu l^{5}}h^{'}(x)+\cr\nonumber\\
&&+\frac{r_{h}^{4}m^{2}}{2l^{6}}\big(2+\frac{1}{m^{2}l^{2}}\big)h(x)=\frac{r_{h}^{4}m^{2}}{2cl^{6}}\big(\kappa lEe^{2\pi t_{w}/\beta}\big)\delta(x),
\end{eqnarray}
we can write the above differential equation in this form:
\begin{eqnarray}
&&\big(\partial_{x}^{2}-b_{2}^{2}\big)\big(\partial_{x}^{2}-a\partial_{x}-b_{1}^{2}\big)h(x)=\xi\delta(x),   \,\,\,\,\      \,\,\,\,\     b_{1}^{2}= \frac{r_{h}^{2}}{2l^{4}}\big(1+2m^{2}l^{2}\big),\cr\nonumber\\
&&a=\frac{r_{h}m^{2}}{l\mu},     \,\,\,\,\     \,\,\,\,\    b_{2}^{2}= \frac{r_{h}^{2}}{l^{4}},   \,\,\,\,\     \,\,\,\,\   \xi=\frac{r_{h}^{4}m^{2}}{2cl^{6}}\big(\kappa lEe^{2\pi t_{w}/\beta}\big),
\end{eqnarray}
Now lets decompose the above differential equation:
\begin{eqnarray}
q^{''}(x)-b_{2}^{2}q(x)&=&\xi\delta(x),\cr\nonumber\\
h^{''}(x)-ah^{'}(x)-b_{1}^{2}h(x)&=&q(x),
\end{eqnarray}
First equation is similar to first equation of Eq(4.7), therefore we have $ q(x)=-\frac{\xi}{2b_{2}}e^{-b_{2}|x|}, $ if we put $ q(x) $ in the above second equation we find:
\begin{eqnarray}
&&h(x)=\frac{\xi}{4b_{2}\sqrt{a^{2}+4b_{1}^{2}}\big(b_{1}^{2}-b_{2}(a+b_{2})\big)}\bigg[-2\sqrt{a^{2}+4b_{1}^{2}}e^{-b_{2}x}+\cr\nonumber\\
&&+\big(\sqrt{a^{2}+4b_{1}^{2}}+2b_{2}+a\big)e^{-\frac{1}{2}\big(-a+\sqrt{a^{2}+4b_{1}^{2}}\big)x}+\big(\sqrt{a^{2}+4b_{1}^{2}}-2b_{2}-a\big)e^{-\frac{1}{2}\big(-a-\sqrt{a^{2}+4b_{1}^{2}}\big)x}\bigg].\cr\nonumber\\
\end{eqnarray}
Using the expressions for $ a $, $ b_{1} $, $ b_{2} $ and $ \xi $ we can read the scrambling time and the butterfly velocities
as follows:
\begin{eqnarray}
&&t_{*}=\frac{\beta}{2\pi}log\frac{l}{\kappa},  \,\,\,\,\      \,\,\,\,\  v_{B}^{(1)}=\frac{2\pi}{\beta b_{2}}=1,\cr\nonumber\\
&&v_{B}^{(2)}=\frac{2\pi}{\beta \big(\frac{1}{2}(-a+\sqrt{a^{2}+4b_{1}^{2}})\big)}= \frac{2\mu}{m^{2}l}\bigg(\frac{1}{-1+\sqrt{1+\frac{2\mu^{2}}{m^{2}}(2+\frac{1}{m^{2}l^{2}})}}\bigg),\cr\nonumber\\
&&v_{B}^{(3)}=\frac{2\pi}{\beta \big(\frac{1}{2}(-a-\sqrt{a^{2}+4b_{1}^{2}})\big)}= \frac{2\mu}{m^{2}l}\bigg(\frac{-1}{1+\sqrt{1+\frac{2\mu^{2}}{m^{2}}(2+\frac{1}{m^{2}l^{2}})}}\bigg), \,\,\  \,\,\  \frac{2\pi}{\beta}=\frac{r_{h}}{l^{2}}.
\end{eqnarray}
Note that for $ \mu^{2}>0 $ and $ m^{2}>0 $ in a ghost free regime $ v_{B}^{(3)} $ is negative which implies moving in backward direction. At the NMG limit of the model, $ \mu \longrightarrow \infty $ we have:
\begin{eqnarray}
v_{B}^{(2)}=\frac{1}{\sqrt{m^{2}l^{2}+\frac{1}{2}}},      \,\,\,\,\,\           \,\,\,\,\,\        v_{B}^{(3)}=-\frac{1}{\sqrt{m^{2}l^{2}+\frac{1}{2}}},
\end{eqnarray}
note that $ v_{B}^{(2)} $ is exactly one of the butterfly velocities in NMG, and also in TMG limit $ m^{2}\longrightarrow \infty $ with finite $ \mu $ one can see:
\begin{eqnarray}
v_{B}^{(2)}=\frac{1}{\mu l},   \,\,\,\,\,\           \,\,\,\,\,\    v_{B}^{(3)}=0,
\end{eqnarray}
here $ v_{B}^{(2)} $ is one of the butterfly velocities in TMG which is in agreement with result of \cite{Alishahiha:2016cjk} for TMG Eq(3.12). In addition there is a critical line in parameter space of GMG at $ \frac{1}{2m^{2}l^{2}}+\frac{1}{\mu l}=1 $,\cite{Liu:2009pha} which in TMG limit, $ m^{2}\longrightarrow \infty $ is the critical point of TMG, $ \mu l=1 $ and in NMG limit, $ \mu \longrightarrow \infty $ is the critical point of NMG, $ m^{2}l^{2}=\frac{1}{2} $. One can see at the critical line of GMG we have,
\begin{eqnarray}
 v_{B}^{(2)}=1,        \,\,\,\,\,\           \,\,\,\,\,\     v_{B}^{(3)}=-\frac{\mu l-1}{\mu l-\frac{1}{2}},
\end{eqnarray}
note that $ v_{B}^{(2)} $ is the butterfly velocity of the 3D Einstein Gravity, and also in NMG limit, $ \mu\longrightarrow \infty $, $ v_{B}^{(3)}=-1 $, which is the butterfly velocity of the 3D Einstein Gravity with negative sign, and is in agreement with Eq(4.21) at the critical point of NMG, $ m^{2}l^{2}=\frac{1}{2} $. In addition one can see at the critical point of TMG, $ \mu l=1 $ we have $ v_{B}^{(3)}=0 $ which is in agreement with Eq(4.22). Note that in TMG limit Eq(4.22) for $ \mu l=-1 $ we have $  v_{B}^{(2)}=-1 $ which is the butterfly velocity of the 3D Einstein Gravity  with negative sign, maybe it means moving in backward direction,  and also here there is an interesting point; maybe the negative butterfly velocities imply some instabilities in the dual theory, these instabilities maybe lead to a phase transition\footnote{I would like to thank the referee for her/his comment on this point.}. In \cite{Ling:2016ibq}, \cite{Ling:2017jik} the authors proposed that the butterfly velocity $ v_{B} $,  can be used to diagnose quantum phase transition (QPT) in holographic theories. They provided evidences for this proposal with a holographic model exhibiting metal-insulator transitions (MIT), in which the derivatives of $ v_{B} $, with respect to system parameters characterizes quantum critical points (QCP) with local extremes in zero temperature limit\cite{Ling:2016ibq}.\\
\\
We can consider $ \mu l=-1 $ as the other critical point of theory, therefor the other critical line of GMG is, $ \frac{1}{2m^{2}l^{2}}-\frac{1}{\mu l}=1 $. We know from the dual 2D CFT of TMG\cite{Li:2008dq}:
\begin{eqnarray}
c_{L}=\frac{3l}{2G}\Big(1-\frac{1}{\mu l}\Big),        \,\,\,\,\,\           \,\,\,\,\,\      c_{R}=\frac{3l}{2G}\Big(1+\frac{1}{\mu l}\Big),       
\end{eqnarray}
one can see at two critical points of TMG, $ \mu l=1 $ and $ \mu l=-1 $ we have two different chiral modes, right-moving and left-moving respectively as follows,
\begin{eqnarray}
\mu l=1, \,\,\,\,\,\           \,\,\,\,\,\   c_{L}=0,  \,\,\,\,\,\           \,\,\,\,\,\  c_{R}=\frac{3l}{G},\cr\nonumber\\
\mu l=-1, \,\,\,\,\,\           \,\,\,\,\,\   c_{L}=\frac{3l}{G},  \,\,\,\,\,\           \,\,\,\,\,\  c_{R}=0.
\end{eqnarray}
And also we know that changing the sign of topological mass in TMG, $ \mu\longrightarrow -\mu $, is equivalent to acting parity operator on the theory and going from left-moving mode to right-moving mode and vise versa.\\
 Now lets consider the other critical line of GMG at $ \frac{1}{2m^{2}l^{2}}-\frac{1}{\mu l}=1 $, at this line the butterfly velocities are:
\begin{eqnarray}
v_{B}^{(3)}=-1,        \,\,\,\,\,\           \,\,\,\,\,\     v_{B}^{(2)}=\frac{\mu l+1}{\mu l+\frac{1}{2}}.
\end{eqnarray}
One can see in NMG
limit, $ \mu\longrightarrow \infty $, $ v_{B}^{(2)}=1 $ which is the butterfly velocity of the 3D Einstein Gravity and is in agreement with Eq(4.21) at the critical point of NMG. In addition one can see at the other critical point of TMG,  $ \mu l=-1 $ we have $ v_{B}^{(2)}=0 $. We can conclude that there is a correspondence between the butterfly velocities and right-left moving degrees of freedom or the central charges of the dual Conformal Field theories.\\
We observed that at both of critical lines at the NMG limit $ \mu\longrightarrow \infty $ we have both of right-moving and left-moving velocities:
\begin{eqnarray}
v_{B}^{(2)}=1     \,\,\,\,\,\           \,\,\,\,\,\     v_{B}^{(3)}=-1,
\end{eqnarray}
note that the New Massive Gravity is a parity-preserving or even parity model\cite{Bergshoeff:2009hq}. But at the TMG limits in critical lines or critical points of TMG, we have just right-moving velocity in one branch and just left-moving velocity in the other branch, in other words, TMG is a parity violating or odd parity
theory:
\begin{eqnarray}
&&\mu l=1,        \,\,\,\,\,\        \,\,\,\,\,\      v_{B}^{(3)}=0,        \,\,\,\,\,\           \,\,\,\,\,\     v_{B}^{(2)}=1,\cr\nonumber\\
&&\mu l=-1,   \,\,\,\         \,\,\,\      v_{B}^{(3)}=-1,        \,\,\,\,\,\           \,\,\,\,\,\     v_{B}^{(2)}=0,
\end{eqnarray}
In other language the theory is chiral at the critical points, $ \mu l=1 $ and $ \mu l=-1 $. This relations are so similar to relations for the central charges of the dual 2D CFT Eq(4.25) at the critical points where theory is chiral. Therefor we observe a correspondence between the butterfly velocities and the central charges of dual 2D CFT at the critical points of TMG.\\
Recently a conjecture has been proposed about the lower bound on diffusion coefficient by "butterfly velocity" \cite{Blake:2016wvh},\cite{Blake:2016sud};
\begin{eqnarray}
D \geq \frac{\hbar v_{B}^{2}}{k_{B}T},
\end{eqnarray}
where $ D $ is the diffusion coefficient, $ k_{B} $ is the Boltzmann constant and $ T $ is the temperature. And also in \cite{Kovtun:2008kx} authors studied a universality, which determines the shear viscosity $ \eta $ and “electrical” conductivity $ \sigma $ in terms of the corresponding "central charges" and naturally leads to a conjectured bound on
conductivity in physical systems. And we know the relation between conductivity, charge susceptibility and diffusion coefficient:
\begin{eqnarray}
D=\frac{\sigma}{\chi},
\end{eqnarray}
where $ \chi $ is charge susceptibility. These bounds on conductivity and diffusion coefficient maybe are the evidence of correspondence between the butterfly velocities and the central charges of the dual conformal field theories.

\section{Born-Infeld 3D Gravity}
In this section we study the butterfly effect in the Born-Infeld 3D Gravity\cite{Gullu:2010pc}, \cite{Gullu:2010st}, \cite{Alishahiha:2010iq}, which include $ AdS_{3} $ vacuum as well as solutions with $ AdS_{2}\times S^{1} $ symmetry. The action of the Born-Infeld 3D Gravity is:
\begin{eqnarray}
S_{BI}= -\frac{m^{2}}{4\pi G}\int d^{3}x\sqrt{-g}F(R,K,S),
\end{eqnarray}
where 
\begin{eqnarray}
F(R,K,S)= \sqrt{1+\frac{1}{2m^{2}}\big(R-\frac{1}{2m^{2}}K-\frac{1}{12m^{4}}S\big)}-\big(1+\frac{\Lambda}{2m^{2}}\big),
\end{eqnarray}
with
\begin{eqnarray}
K\equiv R_{\mu\nu}R^{\mu\nu}-\frac{1}{2}R^{2},   \,\,\,\,\,\           \,\,\,\,\,\    S\equiv 8R^{\mu\nu}R_{\mu\alpha}R^{\alpha}~_{\nu}-6RR_{\mu\nu}R^{\mu\nu}+R^{3}.
\end{eqnarray}
Using this form of the action the equations of motion read\cite{Gullu:2010st}
\begin{eqnarray}
-\frac{\kappa}{4m^{2}} T_{\mu\nu}&=& -\frac{1}{2}Fg_{\mu\nu}+(g_{\mu\nu}\Box-\nabla_{\mu}\nabla_{\nu})F_{R}+F_{R}R_{\mu\nu}+\cr\nonumber\\
&&+\frac{1}{m^{2}}\bigg[2\nabla_{\alpha}\nabla_{\mu}(F_{R}R^{\alpha}~_{\nu})-g_{\mu\nu}\nabla_{\alpha}\nabla_{\beta}(F_{R}R^{\alpha\beta})-\Box(F_{R}R_{\mu\nu})-2F_{R}R_{\nu}~^{\alpha}R_{\mu\alpha}+\cr\nonumber\\
&&+g_{\mu\nu}\Box(F_{R}R)-\nabla_{\mu}\nabla_{\nu}(F_{R}R)+F_{R}RR_{\mu\nu}\bigg]-\cr\nonumber\\
&&-\frac{1}{2m^{4}}\bigg[4F_{R}R^{\rho}~_{\mu}R_{\rho\alpha}R^{\alpha}~_{\nu}+2g_{\mu\nu}\nabla_{\alpha}\nabla_{\beta}(F_{R}R^{\beta\rho}R^{\alpha}~_{\rho})+2\Box(F_{R}R_{\nu}~^{\alpha}R_{\mu\alpha})-\cr\nonumber\\
&&-4\nabla_{\alpha}\nabla_{\mu}(F_{R}R_{\nu}~^{\rho}R^{\alpha}~_{\rho})+2\nabla_{\alpha}\nabla_{\mu}(F_{R}RR^{\alpha}~_{\nu})-g_{\mu\nu}\nabla_{\alpha}\nabla_{\beta}(F_{R}RR^{\alpha\beta})-\cr\nonumber\\
&&-\Box(F_{R}RR_{\mu\nu})-2F_{R}RR_{\nu}~^{\rho}R_{\mu\rho}-g_{\mu\nu}\Box(F_{R}R_{\alpha\beta}R^{\alpha\beta})+\nabla_{\mu}\nabla_{\nu}(F_{R}R_{\alpha\beta}R^{\alpha\beta})-\cr\nonumber\\
&&-F_{R}R_{\alpha\beta}R^{\alpha\beta}R_{\mu\nu}+\frac{1}{2}g_{\mu\nu}\Box(F_{R}R^{2})-\frac{1}{2}\nabla_{\mu}\nabla_{\nu}(F_{R}R^{2})+\frac{1}{2}F_{R}R^{2}R_{\mu\nu}\bigg],
\end{eqnarray}
where
\begin{eqnarray}
F_{R}=\frac{\partial F}{\partial R}=\frac{1}{4m^{2}}\bigg[F+\Big(1+\frac{\Lambda}{2m^{2}}\Big)\bigg]^{-1}.
\end{eqnarray}
The equations of motion of the Born-Infeld 3D Gravity admit this asymptotically AdS black hole solution Eq(2.4) with:
\begin{eqnarray}
\Lambda=-2m^{2}\bigg(1-\sqrt{1-\frac{1}{m^{2}l^{2}}}\bigg).
\end{eqnarray}
Plugging the ansatz in the Kruskal coordinate Eq(2.10) into equations of motion Eq(5.4), near horizon at the leading order one finds a forth order differential equation for $ h(x) $:
\begin{eqnarray}
&&\frac{d^{4}h(x)}{dx^{4}}-\frac{r_{h}^{2}}{l^{4}}(4+m^{2}l^{2})h^{''}(x)+\frac{r_{h}^{4}}{l^{8}}\frac{(3m^{2}l^{2}+1)(m^{2}l^{2}+1)}{m_{2}l^{2}-1}h(x)=\cr\nonumber\\
&&=-\frac{r_{h}^4m}{2cl^{7}}\sqrt{m^{2}l^{2}-1}\big(\kappa lEe^{2\pi t_{w}/\beta}\big).
\end{eqnarray}
One can write the above differential equation in this form:
\begin{eqnarray}
&&\big(\partial_{x}^{2}-b_{1}^{2}\big)\big(\partial_{x}^{2}-b_{2}^{2}\big)h(x)=\xi\delta(x),   \,\,\,\,\      \,\,\,\,\     b_{1}^{2}= \frac{r_{h}^{2}}{2l^{4}}\bigg[4+m^{2}l^{2}+\sqrt{\frac{m^{6}l^{6}-5m^{4}l^{4}-8m^{2}l^{2}-20}{m^{2}l^{2}-1}}\bigg],\cr\nonumber\\
&&b_{2}^{2}= \frac{r_{h}^{2}}{2l^{4}}\bigg[4+m^{2}l^{2}-\sqrt{\frac{m^{6}l^{6}-5m^{4}l^{4}-8m^{2}l^{2}-20}{m^{2}l^{2}-1}}\bigg], \,\,\,\ \,\,\,\   \xi=-\frac{r_{h}^{4}m}{2cl^{7}}\sqrt{m^{2}l^{2}-1}\big(\kappa lEe^{2\pi t_{w}/\beta}\big),\cr\nonumber\\
\end{eqnarray}
If we decompose the above differential equations are similar to Eq(4.7) therefore the solution of $ h(x) $ is exactly the Eq(4.9):
\begin{eqnarray}
h(x)= \frac{\xi}{2b_{1}b_{2}(b_{1}^{2}-b_{2}^{2})}\bigg(b_{1}e^{-b_{2}x}-b_{2}e^{-b_{1}x}\bigg).
\end{eqnarray}
Using the expressions for $ b_{1} $ ,  $ b_{2} $ and $ \xi $ we can read the scrambling time and the butterfly velocities as follows:
\begin{eqnarray}
&&t_{*}=\frac{\beta}{2\pi}log\frac{l}{\kappa},  \,\,\,\,\      \,\,\,\,\  v_{B}^{(1)}=\frac{2\pi}{\beta b_{1}}=\sqrt{\frac{2\sqrt{m^{2}l^{2}-1}}{(4+m^{2}l^{2})\sqrt{m^{2}l^{2}-1}+\sqrt{m^{6}l^{6}-5m^{4}l^{4}-8m^{2}l^{2}-20}}},\cr\nonumber\\
&&v_{B}^{(2)}=\frac{2\pi}{\beta b_{2}}=\sqrt{\frac{2\sqrt{m^{2}l^{2}-1}}{(4+m^{2}l^{2})\sqrt{m^{2}l^{2}-1}-\sqrt{m^{6}l^{6}-5m^{4}l^{4}-8m^{2}l^{2}-20}}},  \,\,\,\,\        \,\,\,\,\  \frac{2\pi}{\beta}=\frac{r_{h}}{l^{2}}.\cr\nonumber\\
\end{eqnarray}
Now lets consider the critical point of the Born-Infeld 3D Gravity, $ m^{2}l^{2}=1 $ where the model has logarithmic wave solutions\cite{Alishahiha:2010iq}. One can see at the critical point, the above two velocities degenerate and are equal to zero
\begin{eqnarray}
m^{2}l^{2}=1,   \,\,\,\,\,\,\           \,\,\,\,\,\,\   v_{B}^{(1)}=v_{B}^{(2)}=0.
\end{eqnarray}
In\cite{Qaemmaqami:2017bdn} we observed that by adding higher curvature correction to the Einstein Gravity, the butterfly velocity decreases at the critical point. It's interesting that the butterfly velocities in the Born-Infeld 3D Gravity vanish at the critical point, it's important to note that the Born-Infeld 3D Gravity has infinite higher derivative in level of the action because the square root form of the action.\\
And also it has worth to note that at the critical point of the Born-Infeld Gravity, both of the central charges of the dual 2D CFT vanish\cite{Gullu:2010st},\cite{Alishahiha:2010iq}:
\begin{eqnarray}
m^{2}l^{2}=1,        \,\,\,\,\,\,\           \,\,\,\,\,\,\    c_{L}=c_{R}= \frac{3l}{2G}\sqrt{1-\frac{1}{m^{2}l^{2}}}=0.
\end{eqnarray}
Maybe we can say it's an other evidence for correspondence between the butterfly velocities and the central charges of the dual 2D CFT.
\section{New Bi-Gravity}
The New Bi-Gravity is a recently proposed 3D Gravity model for resolving bulk-boundary clash in the New Massive Gravity\cite{Akhavan:2016hju},\cite{Akhavan:2016xmj}, if we consider the NMG action by using an auxiliary field, $ f_{\mu\nu} $ then promote the auxiliary field to dynamical field. The New Bi-Gravity(NBG) action is:
\begin{eqnarray}
S_{NBG}&=&\frac{1}{16\pi G}\int d^{3}x\sqrt{-g}\bigg(\sigma R[g]+f_{\mu\nu}\mathcal{G}^{\mu\nu}[g]+\frac{1}{4}m^{2}(\tilde{f}^{\mu\nu}f_{\mu\nu}-\tilde{f}^{2})-2\Lambda_{g}\bigg)+\cr\nonumber\\
&&+\frac{1}{16\pi\tilde{G}}\int d^{3}x\sqrt{-f}\bigg(R[f]-2\Lambda_{f}\bigg),
\end{eqnarray}
where $ \Lambda_{f} $ is a new cosmological constant, $ \tilde{G} $ is Newton constant of the new metric, $ R[g] $ and $ R[f] $ are Ricci scalars constructed from  $ g_{\mu\nu} $ and  $ f_{\mu\nu} $ respectively. $ \mathcal{G}_{\mu\nu} $ is the Einstein tensor of the metric  $ g_{\mu\nu} $. Note that all indices are raised by $ g^{\mu\nu} $ except those in the definition of Ricci scalar $ R[f] $ which are raised by the inverse metric $ f^{\mu\nu} $.\\
By varying the above NBG action with respect to the metrics $ g_{\mu\nu} $ and  $ f_{\mu\nu} $ one can find the equations of motion:
\begin{eqnarray}
&&\mathcal{G}[g]_{\mu\nu}+\Lambda_{g}g_{\mu\nu}+\frac{m^{2}}{2}\bigg[\tilde{f}_{\mu}~^{\rho}f_{\nu\rho}-\tilde{f}f_{\mu\nu}-\frac{1}{4}g_{\mu\nu}(\tilde{f}^{\rho\sigma}f_{\rho\sigma}-\tilde{f}^{2})\bigg]+2\tilde{f}_{(\mu}~^{\rho}\mathcal{G}[g]_{\nu)\rho}+\cr\nonumber\\
&&+\frac{1}{2}f_{\mu\nu}R[g]-\frac{1}{2}\tilde{f}R_{\mu\nu}[g]-\frac{1}{2}g_{\mu\nu}f_{\rho\sigma}\mathcal{G}[g]^{\rho\sigma}+\frac{1}{2}\bigg[\nabla^{2}[g]f_{\mu\nu}-2\nabla[g]^{\rho}\nabla[g]_{(\mu}f_{\nu)\rho}+\cr\nonumber\\
&&+\nabla[g]_{\mu}\nabla[g]_{\nu}\tilde{f}+\big(\nabla[g]^{\rho}\nabla[g]^{\sigma}f_{\rho\sigma}-\nabla^{2}[g]\tilde{f}\big)g_{\mu\nu}\bigg]=\kappa T_{\mu\nu},
\end{eqnarray}
\begin{eqnarray}
\mathcal{G}[f]_{\mu\nu}+\Lambda_{f}f_{\mu\nu}-\frac{1}{k}\sqrt{\frac{g}{f}}\bigg[f_{\alpha\mu}f_{\beta\nu}\mathcal{G}[g]^{\alpha\beta}+\frac{1}{2}m^{2}(g^{\sigma\alpha}g^{\tau\beta}-g^{\sigma\tau}g^{\alpha\beta})(f_{\sigma\tau}f_{\alpha\mu}f_{\beta\nu})\bigg]=\kappa T_{\mu\nu},
\end{eqnarray}
where $ \mathcal{G}[f]_{\mu\nu} $ is the Einstein tensor of the metric $ f_{\mu\nu} $ and $ k=\frac{G}{\tilde{G}} $ is the relative strength of two Newton constants associated with two metrics.\\
The equations of motion of New  Bi-Gravity admit this asymptotically AdS black hole solution:
\begin{eqnarray}
ds_{g}^{2}=-f(r)dt^{2}+\frac{dr^{2}}{f(r)}+\frac{r^{2}}{l^{2}}dx^{2},    \,\,\,\,\    \,\,\,\,\    f(r)=\frac{r^{2}}{l^{2}}\bigg(1-\frac{r_{h}^{2}}{r^{2}}\bigg), \,\,\,\,\      \,\,\,\,\    ds_{f}^{2}=\gamma ds_{g}^{2},
\end{eqnarray}
with
\begin{eqnarray}
4-\gamma^{2}l_{g}^{2}m^{2}+2\gamma+4l_{g}^{2}\Lambda_{g}=0,   \,\,\,\,\,\,\         \,\,\,\,\,\,\   1-m^{2}l_{f}^{2}-k\frac{l_{g}}{l_{f}}(1+\Lambda_{f}l_{f}^{2})=0.
\end{eqnarray}
For solving the equations of motion near horizon to find the shock wave solution, we consider these ansatz in the Kruskal coordinate for back-reacted geometry
\begin{eqnarray}
ds_{g}^{2}&=&2A(UV)dUdV+B(UV)dx^{2}-2A(UV)h(x)\delta(U)dU^{2},\cr\nonumber\\
ds_{f}^{2}&=&2 A(UV)dUdV+ B(UV)dx^{2}-2 A(UV)\rho(x)\delta(U)dU^{2},
\end{eqnarray}
note that here we take $ \gamma=1 $ in other words we take same background for $ g_{\mu\nu} $ and $ f_{\mu\nu} $ but the perturbations around background are different\cite{Akhavan:2016hju} by $ h(x) $ and $ \rho(x) $ functions. Plugging the ansatz in the Kruskal coordinate Eq(6.6) into the equations of motion Eq(6.2) and Eq(6.3), near horizon at the leading order one finds two coupled forth order differential equation for $ h(x) $ and $ \rho(x) $: 
\begin{eqnarray}
&&2\rho^{''}(x)-5h^{''}(x)-2\frac{r_{h}^{2}}{l^{4}}(l^{2}m^{2}-1)\rho(x)+\frac{r_{h}^{2}}{l^{4}}(2l^{2}m^{2}+1)h(x)=\xi \delta(x),\cr\nonumber\\
&&k\rho^{''}(x)-h^{''}(x)-\frac{r_{h}^{2}}{l^{4}}(l^{2}m^{2}+k-2)\rho(x)+\frac{r_{h}^{2}}{l^{4}}(l^{2}m^{2}-1)h(x)=-\frac{k}{2}\xi \delta(x),
\end{eqnarray}
where $ \xi= \frac{r_{h}^{2}}{cl^{4}}\big(\kappa lEe^{2\pi t_{w}/\beta}\big) $. 
The solutions of the above coupled differential equations are:
\begin{eqnarray}
h(x)&=&\xi\bigg[\frac{-(1+k)}{2r_{h}l^{2}(1+2k)}e^{-\frac{r_{h}}{l^{2}}x}+\cr\nonumber\\
&&+\frac{k^{2}+k-2}{2r_{h}l^{2}(1+2k)\sqrt{5k-2}\sqrt{l^{2}m^{2}(1+2k)+k-4}} e^{-\frac{r_{h}}{l^{2}}\sqrt{\frac{l^{2}m^{2}(1+2k)+k-4}{5k-2}}x}\bigg],\cr\nonumber\\
\rho(x)&=&\xi\bigg[\frac{-(1+k)}{2r_{h}l^{2}(1+2k)}e^{-\frac{r_{h}}{l^{2}}x}-\cr\nonumber\\
&&-\frac{3(k+2)}{2r_{h}l^{2}(1+2k)\sqrt{5k-2}\sqrt{l^{2}m^{2}(1+2k)+k-4}}e^{-\frac{r_{h}}{l^{2}}\sqrt{\frac{l^{2}m^{2}(1+2k)+k-4}{5k-2}}x}\bigg].
\end{eqnarray}
From the above expressions, one can read the scrambling time and the butterfly velocities as follows:
\begin{eqnarray}
&&t_{*}=\frac{\beta}{2\pi}log\frac{l}{\kappa},  \,\,\,\,\      \,\,\,\,\  v_{B}^{(1)}=\frac{2\pi}{\beta \big(\frac{r_{h}}{l^{2}}\big)}=1,\cr\nonumber\\  &&v_{B}^{(2)}=\frac{2\pi}{\beta \bigg(\frac{r_{h}}{l^{2}}\sqrt{\frac{l^{2}m^{2}(1+2k)+k-4}{5k-2}}\bigg)}=\sqrt{\frac{5k-2}{l^{2}m^{2}(1+2k)+k-4}}, \,\,\,\,\,\      \,\,\,\,\,\    \frac{2\pi}{\beta}=\frac{r_{h}}{l^{2}}.
\end{eqnarray}
Now lets consider the $ k=1 $ case which happens when two Newton constants are equal,
\begin{eqnarray}
k=1,     \,\,\,\,\,\,\          \,\,\,\,\,\,\     v_{B}^{(2)}= \frac{1}{\sqrt{m^{2}l^{2}-1}}.
\end{eqnarray}
For respecting the causality, the butterfly velocity must be equal or less than velocity of light, $ v^{(2)}_{B}\leq 1 $ therefor we have:
\begin{eqnarray}
m^{2}l^{2}\geq 2,
\end{eqnarray}
for $ m^{2}l^{2}=2 $ the butterfly velocity is equal to velocity of light, $ v^{(2)}_{B}= 1 $, which is the butterfly velocity of the 3D Einstein Gravity.\\
Finally it's interesting to consider the logarithmic solutions limit of the New Bi-Gravity\cite{Akhavan:2016hju}.
\begin{eqnarray}
1-\frac{\gamma}{2}+k\sqrt{\gamma}=0,
\end{eqnarray}
here we take $ \gamma=1 $, then $ k=-\frac{1}{2} $, therefore we have:
\begin{eqnarray}
k=-\frac{1}{2},       \,\,\,\,\,\,\          \,\,\,\,\,\,\       v_{B}^{(1)}=v_{B}^{(2)}=1,
\end{eqnarray}
this situation is similar to critical points of TMG, MMG and NMG where the models have logarithmic solutions and the two butterfly velocities degenerate into one and it is equal to the butterfly velocity of the 3D Einstein Gravity, which is the velocity of light.

\section{Conclusions}
In this paper we study some of 3-dimensional Gravity models, we calculate the butterfly velocities of these models and also we consider critical points and different limits in some of them.
In section 2 we study the butterfly effect in the 3D Einstein Gravity by considering the shock wave in the Kruskal coordinate near the horizon of the AdS black hole, and we find that the butterfly velocity of the 3D Einstein Gravity is equal to the velocity of light, which is in agreement with\cite{Shenker:2013pqa} in $ D=3 $. Although in the 3D Einstein Gravity there is no propagating degree of freedom or graviton in the bulk, due to boundary degrees of freedom or boundary gravitons the butterfly velocity is non-zero.\\
\\
In section 3 we study the butterfly effect of the Minimal Massive 3D Gravity, we consider the TMG limit of the model which was in agreement with results of\cite{Alishahiha:2016cjk} and we study the critical point of the model and we observed that the two butterfly velocities degenerate at the critical point and are equal to the butterfly velocity of the 3D Einstein Gravity.\\
\\
In section 4 we first review the butterfly effect in the New Massive Gravity  by details and consider the critical point of the model where the two butterfly velocities degenerate and are equal to the butterfly velocity of the 3D Einstein Gravity, then we study the butterfly effect in the Generalized Massive Gravity and we find three butterfly velocities for this theory. Then we consider TMG and NMG limits of theory and critical lines and critical points of the model and we observed that there is a correspondence between the butterfly velocities and right-left moving degrees of freedom or the central charges of the dual 2-dimensional Conformal Field Theory.\\
\\
In section 5 we study the butterfly effect in the Born-Infeld 3D Gravity and we find that at the critical point of the model, the two butterfly velocities degenerate and are equal to zero. It's interesting that the butterfly velocities in the Born-Infeld 3D Gravity vanish at the critical point, we know that the Born-Infeld 3D Gravity has infinite higher derivative in level of the action because of square root form of the action and also in\cite{Qaemmaqami:2017bdn} we observed that by adding higher curvature correction to the Einstein Gravity, the butterfly velocity decreases at the critical point. And also both of the central charges of the dual 2D CFT vanish at the critical point of the model, maybe it's an other evidence for correspondence between the butterfly velocities and the central charges of the dual 2D CFT.\\
\\
In section 6 we study the butterfly effect in the New Bi-Gravity model and we find a causality bound in the parameter space of the model and also we consider the logarithmic solutions limit of the New Bi-Gravity and we observed that in this limit the two butterfly velocities degenerate into one which is equal to the butterfly velocity of the 3D Einstein Gravity.\\
In following it is so important and also interesting to study  the butterfly effect in the dual 2D Conformal Field Theories\cite{Roberts:2014ifa},\cite{Perlmutter:2016pkf},\cite{Turiaci:2016cvo} of these models and rederive the obtained results by CFT calculations.\\

\section*{Acknowledgment}
I am grateful to Ali Naseh for useful discussions and comments. I would like to thank referee for her/ his useful comments. I also acknowledge the use of M. Headricks excellent Mathematica package diffgeo. I would like to thank him for his generosity.

%Bibliography

%--------------------------------------------------------------------
\end{document}